\documentclass{article}
\usepackage[utf8]{inputenc}
\usepackage[T1]{fontenc}
\usepackage{amsmath}
\usepackage{graphicx}
\usepackage{setspace}
\usepackage{epstopdf}

\makeatletter

\title{Nanosecond Electron Holography by Interference Gating}

\author{\underline{Tolga Wagner}, Tore Niermann, Felix Urban, Michael Lehmann}

\date{ }

\begin{document}

\maketitle

\begin{center}
    Technische Universität Berlin, Institut für Optik und Atomare Physik, Straße des 17. Juni 135, 10623 Berlin
\end{center}


\section*{Abstract}

The interference gating is a novel method for robust time-resolved electron holographic measurements by directly switching the interference. Here, a new arrangement is presented in which a biprism in the condenser aperture as a fast electric phase shifter is used to control the interference pattern. High-frequency stimulation of the electric phase shifter in the gigahertz range are performed and observed via electron holography, proving the feasibility of interference gating in the upper picosecond range. Despite the bandwidth limitation of 180~MHz of the current signal generator, a time resolution of 100 nanoseconds is achieved through forward correction of the control signal. With this time resolution, it is already possible to measure the transient response of the biasing holder system. Our method paves the way towards a closer look on fast dynamic processes with high temporal and spatial resolution.

\section*{Highlights}

\begin{itemize}
    \item biprism in condenser aperture plane is used as fast electric phase shifter
    \item achieving 100 nanosecond time resolution with interference gating
    \item measuring transient response of biasing holder system
    \item proving feasibility of interference gating with time resolution in the upper picosecond range
\end{itemize}

\newpage
\section*{Introduction}

With a growing interest in decoding and describing dynamic processes down to nature's smallest components, time-resolved electron microscopy is also becoming increasingly important. Pump probe methods for the investigation of these processes on the smallest time scales are more and more being employed in transmission electron microscopy (TEM) [\ref{itm:bost1}, \ref{itm:zewail}]. So far, the two most common approaches are the stroboscopic illumination with an optically pulsed electron emitter (down to the femtosecond range [\ref{itm:feist}]) or the use of an ultrafast electron detector with time resolutions in the nanosecond range [\ref{itm:pitters}]. Great efforts are made to combine time resolutions down to the femtosecond range with the atomic spatial resolution of a TEM. This makes, for predictable and repeatable processes, the molecular movie, as one of the highest goals in dynamic imaging, seem to be within reach.\\
\\
For electron holography (EH), however, there are only a few examples in which time-resolved experiments were carried out [\ref{itm:zewail}, \ref{itm:kralle}, \ref{itm:kralle2}, \ref{itm:kralle3}, \ref{itm:soma}, \ref{itm:flor}]. With the advantage of making amplitude and phase information separately accessible and thus being able to measure projected potential distributions inside and around a specimen, EH offers the best possibilities for investigations of the electrical nature of nano devices [\ref{itm:twitchett1}, \ref{itm:twitchett2}]. Especially with continuously new technologies for in situ stimulation of specimen (e.g. electrically [\ref{itm:electric1}, \ref{itm:electric2}]), phase retrieving methods like EH as a tool for investigation of devices in-operando are gaining interest [\ref{itm:coop1}, \ref{itm:coop2}]. The same applies to magnetic samples too.\\
\\
Consequently, a promising alternative for time-resolved TEM measurements is interference gating (iGate) [\ref{itm:batman}]. It is a simple method that enables robust time-resolved electron holographic measurements by switching the interference on and off. In the first attempt the achieved time resolution was only in the range of a few 100 $\mu$s and acquisitions suffered from artifacts. But one benefit is that, unlike in the case of stroboscopic illumination, the specimen remains illuminated throughout the entire process. Although the total electron dose increases as a result, static irradiation reduces additional parasitic effects such as charge fluctuations or thermal transients for dose insensitive samples.\\
\\
Here we present new concepts to implement iGate to a more or less common TEM setup with minimal extra equipment. These allow acquisitions without previously known artifacts and with a temporal resolution of 100 ns. The limits and challenges of this implementation with regard to the highest temporal resolution are investigated and discussed. As a first application of iGate the transient response of a MEMS based biasing holder system is measured by investigating the dynamic electric potential of a coplanar micro capacitor.

\section*{Interference gating}

\begin{figure}[h]
\begin{centering}
\includegraphics[width=1\columnwidth]{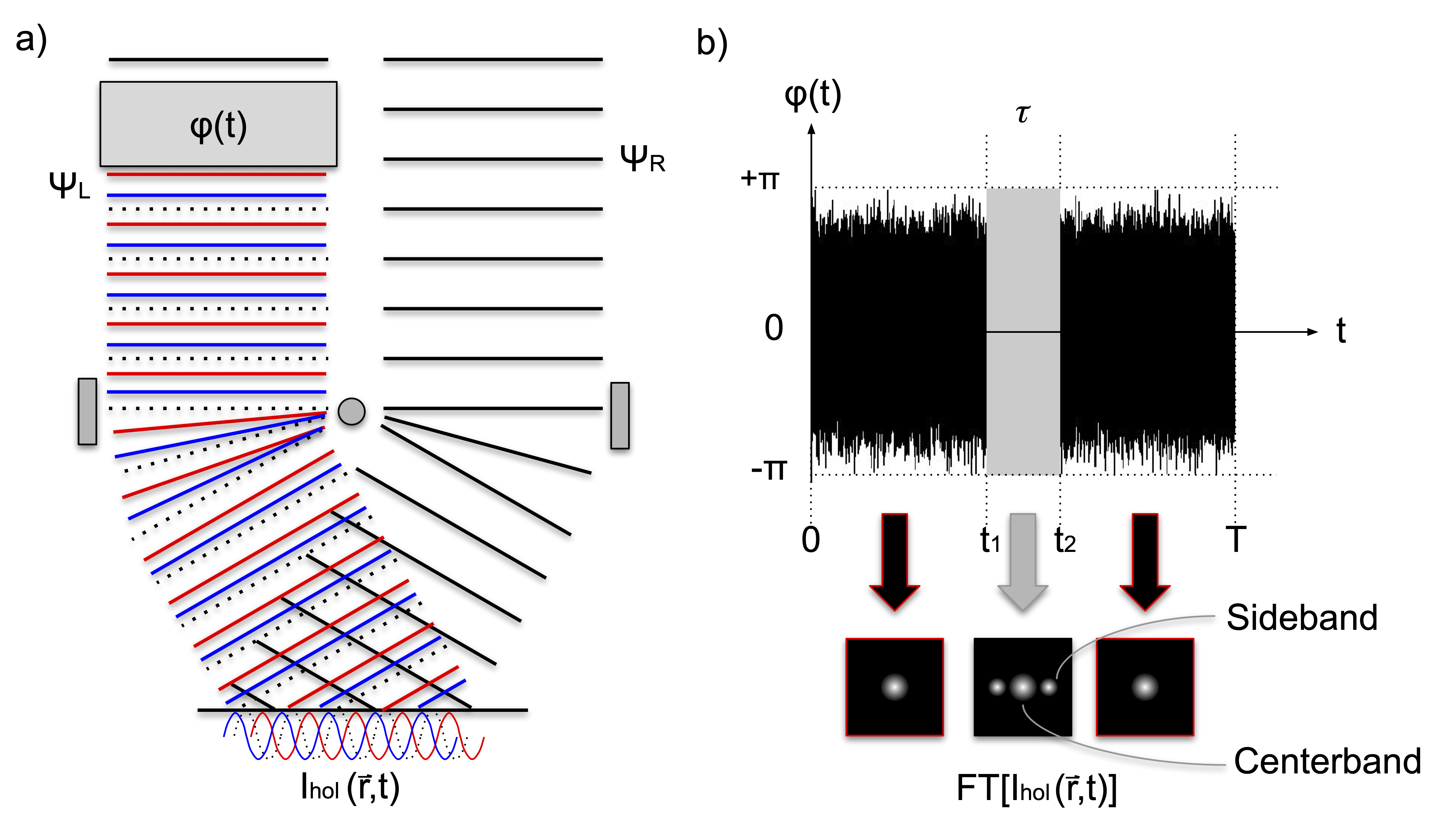}
\par\end{centering}
\caption{a) Dynamic phase shift $\varphi(t)$ producing time-dependent electron hologram $I_{hol}(\vec{r}, t)$ (red and blue lines represent exemplary wave fronts for different times $t$). b) Time-dependent phase sequence for generating time-resolved electron holograms with gating length $\tau = t_2 - t_1$, period T and respective Fourier transformations of $I_{hol}(\vec{r}, t)$ (disturbed case: only centerband, undisturbed: also sidebands).\label{fig:concept}}
\end{figure}

The interference gating enables time-resolved holographic measurements of periodic processes. The basic idea is a synchronized destruction of the interference pattern for most of the time during an interferometric measurement by exposing the setup to controllable instabilities. Only for a defined period of time the setup is kept stable so that the interference pattern can form. The conventional holographic reconstruction process acts as a temporal filter that only retains the information of the undisturbed interferogram inside this period. Although this method can be applied to all interferometric measurements, the following description is limited to an application to off-axis electron holography.\\
\\
An electron hologram is formed by the interference between two wave fronts, object wave $\psi_{L}$ and reference wave $\psi_{R}$. For the sake of simplicity, we here assume that (1) both waves are partially coherent, (2) have planar wave fronts with an amplitude of 1, (3) have a mutual constant phase difference of $\varphi$, and (4) overlap with an interference angle $\beta = \lambda|\vec{q_{c}}|/2$ given by the wavelength $\lambda$ and the carrier frequency $\vec{q_{c}}$ as shown in Fig. \ref{fig:concept} a). 

\noindent
The constant phase shift $\varphi$ is due to the characteristics of the measurement setup, since an empty ray path without an object is assumed for further simplification. The interference pattern is given by:

\begin{align}
I_{hol}(\vec{r}) =  2 + 2\mu \cos(2\pi\vec{q_{c}}\cdot\vec{r}+\varphi),\label{eq:hologram}
\end{align}
\\
where the constant $\mu$ subsumes all contrast reducing effects, i.e. degree of coherence between both waves, instrumental instabilities as well as detector response (MTF). According to Michelson's definition, the fringe contrast
\[
V=\frac{\max(I_{hol}(\vec{r}))-\min(I_{hol}(\vec{r}))}{\max(I_{hol}(\vec{r}))+\min(I_{hol}(\vec{r}))}
\]
is given by $\mu$.\\
\\
The electron holographic reconstruction, which makes use of Fourier optics (including isolation and centering of one of the sidebands in Fourier space and its inverse Fourier transformation, Fig. \ref{fig:concept} b)), provides amplitude and phase information of the image wave [\ref{itm:lichte}].\\
\\
If the phase relation between both waves varies during the exposure time $T_{exp}$ hence $\varphi\rightarrow\varphi(t)$, the averaged interference pattern becomes:

\begin{align}
I_{hol}(\vec{r})  & = \frac{1}{T_{exp}}\int_{0}^{T_{exp}} I_{hol}(\vec{r},t)\,\mathrm{d}t\nonumber \\& = \frac{1}{T_{exp}}\int_{0}^{T_{exp}} \left[2 + 2\mu \cos(2\pi\vec{q_{c}}\cdot\vec{r}+\varphi(t))\right]\,\mathrm{d}t\nonumber \\
 & =2 + \frac{2\mu}{T_{exp}} \int_{0}^{T_{exp}}\cos(2\pi\vec{q_{c}}\cdot\vec{r}+\varphi(t))\,\mathrm{d}t. \label{eq:dynamichologram1}
\end{align}
\\
For periodic variations of $\varphi(t)$ with a period $T$ as an integer divider of $T_{exp}$ ($T_{exp}=n T$), equation (\ref{eq:dynamichologram1}) can be written as:

\begin{align}
I_{hol}(\vec{r}) & = 2 + \frac{2\mu}{T} \int_{0}^{T}\cos(2\pi\vec{q_{c}}\cdot\vec{r}+\varphi(t))\,\mathrm{d}t. \label{eq:dynamichologram2}
\end{align}
\\
If the time-dependent phase shift $\varphi(t)$, as shown in Fig. \ref{fig:concept} b), becomes zero for a time interval $[t_{1},t_{2}]$ with a length $\tau=t_{2}-t_{1}$ (gate length) and has equally distributed random values between $\pm \pi$ outside the interval, equation (\ref{eq:dynamichologram2}) evolves to

\begin{align}
I_{hol}(\vec{r}) & = 2 + \frac{2\mu}{T} \left[\int_{t_{1}}^{t_{2}}\cos(2\pi\vec{q_{c}}\cdot\vec{r})\,\mathrm{d}t+\int_{T \backslash [t_{1},t_{2}]}\cos(2\pi\vec{q_{c}}\cdot\vec{r}+\varphi(t))\,\mathrm{d}t\right] \nonumber \\
& = 2 + \frac{2\mu}{T} \left[\tau\cos(2\pi\vec{q_{c}}\cdot\vec{r})+\underbrace{cos(2\pi\vec{q_{c}}\cdot\vec{r})\int_{T \backslash [t_{1},t_{2}]}^{}\cos(\varphi(t))\,\mathrm{d}t}_{=0}  -\underbrace{sin(2\pi\vec{q_{c}}\cdot\vec{r})\int_{T \backslash [t_{1},t_{2}]}\sin(\varphi(t))\,\mathrm{d}t}_{=0}\right]. \label{eq:dynamichologram3}
\end{align}
\\
The last two terms in (\ref{eq:dynamichologram3}) evaluate to zero for equally distributed random $\varphi(t)$ between $\pm \pi$, hence no interferometric information is generated in the time range outside $\tau$. Such a phase sequence (gating-signal) acts as a kind of gating or shutter, which switches the interference on and off. Only the gate $\tau$ contributes to the electron hologram:

\begin{align}
I_{hol}(\vec{r},\forall t\in\tau) & = 2+2\mu\frac{\tau}{T} \cos(2\pi\vec{q_{c}}\cdot\vec{r}). \label{eq:gate}
\end{align}
\\
While the gate length $\tau$ determines the time resolution of the measurement, the ratio $\frac{\tau}{T}$ (gate fraction), as shown in (\ref{eq:gate}), reduces the fringe contrast. The signal-to-noise ratio (SNR) of the phase is therefore decreased by this fraction compared to a conventional (static) hologram with an exposure time of $T$. For periodic processes, the impact of these effects can partially be compensated by repeating the gating with a repetition rate $f$ during the exposure and by averaging over several holograms [\ref{itm:tore}].\\
\\
A holographic reconstruction of a hologram acquired as described above over an exposure time $T_{exp}$ provides only averaged amplitude and phase information of the image wave during the time interval $\tau$ (or multitude of it). The information outside this interval is filtered out by the reconstruction process itself.\\
\\
This description refers to the measurement process itself (no specimen or dynamic process involved). An extension of the theory for dynamic amplitudes and phases of a specimen, at least if their changes in time are uncorrelated to the external shift of the phase shifter $\varphi(t)$, provides information on averaged amplitudes and phases of the object wave $\psi_{L}$ from the gate $\tau$. By shifting the gate in time (i.e. changing the gate position) in respect to a periodic process, the whole process can be sampled in a kind of pump-probe experiment. Here the number of independent sampling points is given by the gate fraction $\frac{\tau}{T}$.

\section*{Phase shifter}

\begin{figure}[h]
\begin{centering}
\includegraphics[width=1\columnwidth]{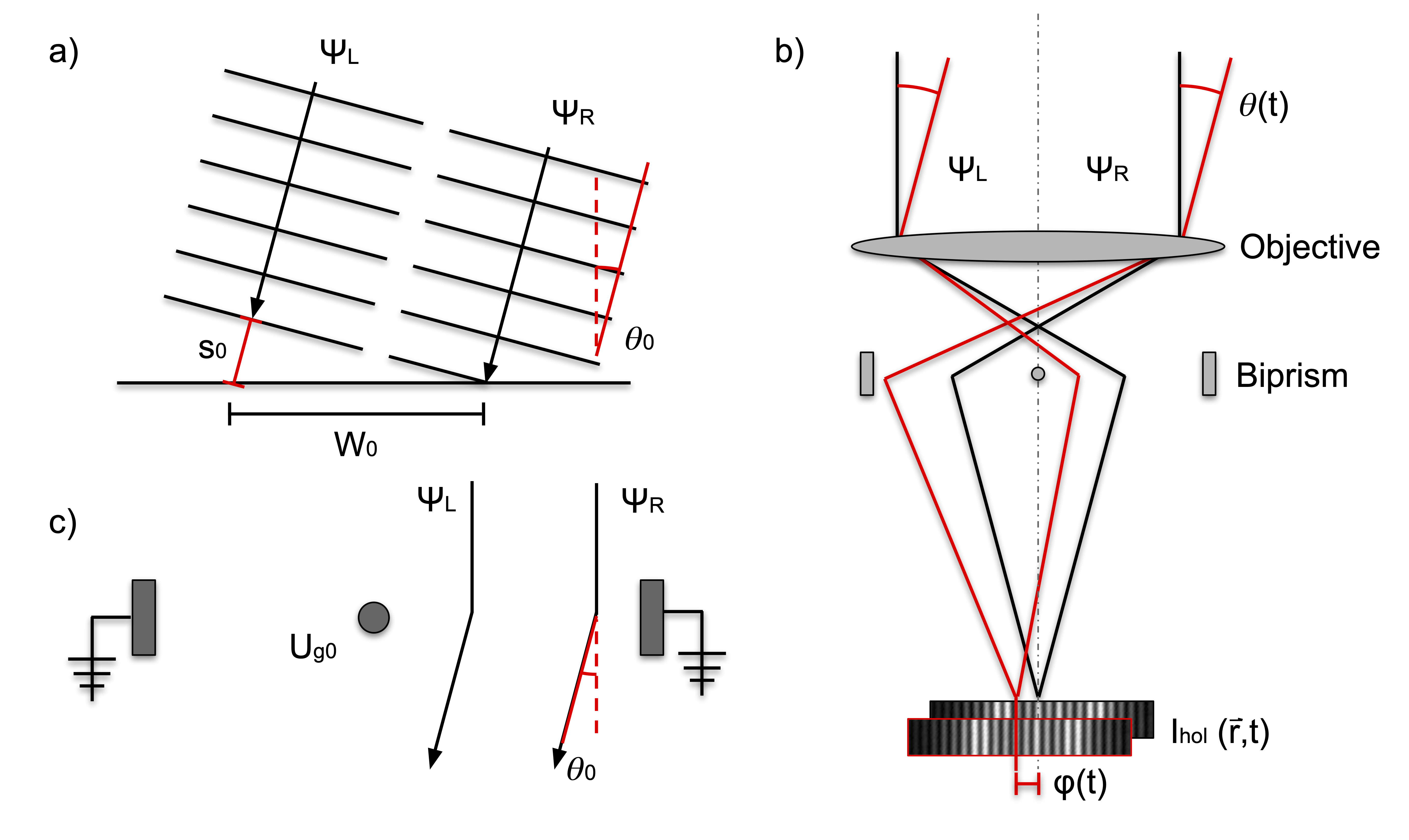}
\par\end{centering}
\caption{a) Wave front of a incident beam tilted by $\theta_{0}$ producing a path difference $S_{0}$. b)  Schematic of experimental realization of the dynamic phase shift $\varphi(t)$ produced by a dynamic beam tilt $\theta(t)$ in an electron holographic setup. Black and red lines indicate the ray path of the normal and tilted electron beam. c) Schematic of a biprism used as deflector by passing $\psi_{L}$ and $\psi_{R}$ between the filament on the electric potential $U_{g0}$ and one grounded counter electrode.\label{fig:concept2}}
\end{figure}

A dynamic phase shifter, capable of producing uniformly distributed random phase shifts from $-\pi$ to $+\pi$ between object wave $\psi_{L}$ and reference wave $\psi_{R}$, can be used to switch the interference. There are several ways to realize such a dynamic phase shifter in a TEM.\\
\\
Originally, a voltage variation of the biprism, which produces the overlap between $\psi_{L}$ and $\psi_{R}$, was used to change the interference angle $\beta$. This causes a time-dependent modulation of both waves, but creates an additional artifact which allows only the outer areas of the hologram to be evaluated in a time-resolved manner [\ref{itm:batman}].\\
\\
Another simple, yet effective way to produce a mutual phase shift between $\psi_{L}$ and $\psi_{R}$ is a slight tilt of the incident beam [\ref{itm:ru1}, \ref{itm:ru2}]. As showed in Fig. \ref{fig:concept2} a) the tilting angle $\theta_{0}$ of the incident beam produces a path difference
\begin{align*}
    S_{0}=W_{0}\tan(\theta_{0})\propto W_{0}\,\theta_{0},
\end{align*}
where $W_{0}$ is the distance between both waves assuming only small tilting angles. For a given de Broglie wavelength $\lambda$, this results in a phase shift of
\begin{align*}
    \varphi_{0}=\frac{2\pi}{\lambda}S_{0}=\frac{2\pi}{\lambda} W_{0}\,\theta_{0}.
\end{align*}
The desired beam tilting is achieved by utilizing the AC beam deflection system of the microscope.\\
\\
Since almost every modern TEM is equipped with such deflectors, time-resolved experiments can be realized without direct intervention in the ray path. These systems, however, are designed for scanning transmission  electron microscopy (STEM) operation and can be switched in the range of a few microseconds, because in STEM-mode the fly-back time is around 100~$\mu s$). During tests of this particular realization, time-resolved electron holograms with a gate fraction of $\frac{\tau}{T} = 0.25$ at a repetition rate of $f = 200$~$kHz$ were realized with a conventional arbitrary waveform generator (GW Instek MFG-2260MRA) as a signal source for the deflectors. This corresponds to a time resolution of 1.25~$\mu s$. However, typical AC deflectors in a TEM (as a magnetic deflection system) are not designed for higher signal frequencies due to their high Impedance, also given by their inductance.\\

\noindent
Since only a slight beam tilt is necessary to disturb the interference, an electrical deflection system can also be used. Providing a low capacity, this may improve the time resolution by several orders of magnitude. However, TEMs are usually not equipped with electric deflectors. But an electron optical biprism (e.g. in the condenser aperture plane) can be used as an electrical deflector if, as shown in Fig. \ref{fig:concept2} c), the whole electron beam is passed between the biprism filament and one counter electrode. When a voltage $U_{g0}$ is applied to the filament, an electric field is generated which is predominantly perpendicular to the electron beam resulting in a beam tilt. Since the deflection angle for small tilts is proportional to the applied voltage $U_{g0}$, the phase shift results approximately in 
\begin{align*}
    \varphi_{0}\propto U_{g0}.
\end{align*}
A dynamic voltage $U_{g}(t)$ applied to the electric deflector (e.g. created by a signal generator) will produce a dynamic phase shift $\varphi(t)$, which can be used to switch the interference for realization of time-resolved electron holography.

\subsection*{Limitations of the dynamic phase shifter}

\begin{figure}[h]
\begin{centering}
\includegraphics[width=1\columnwidth]{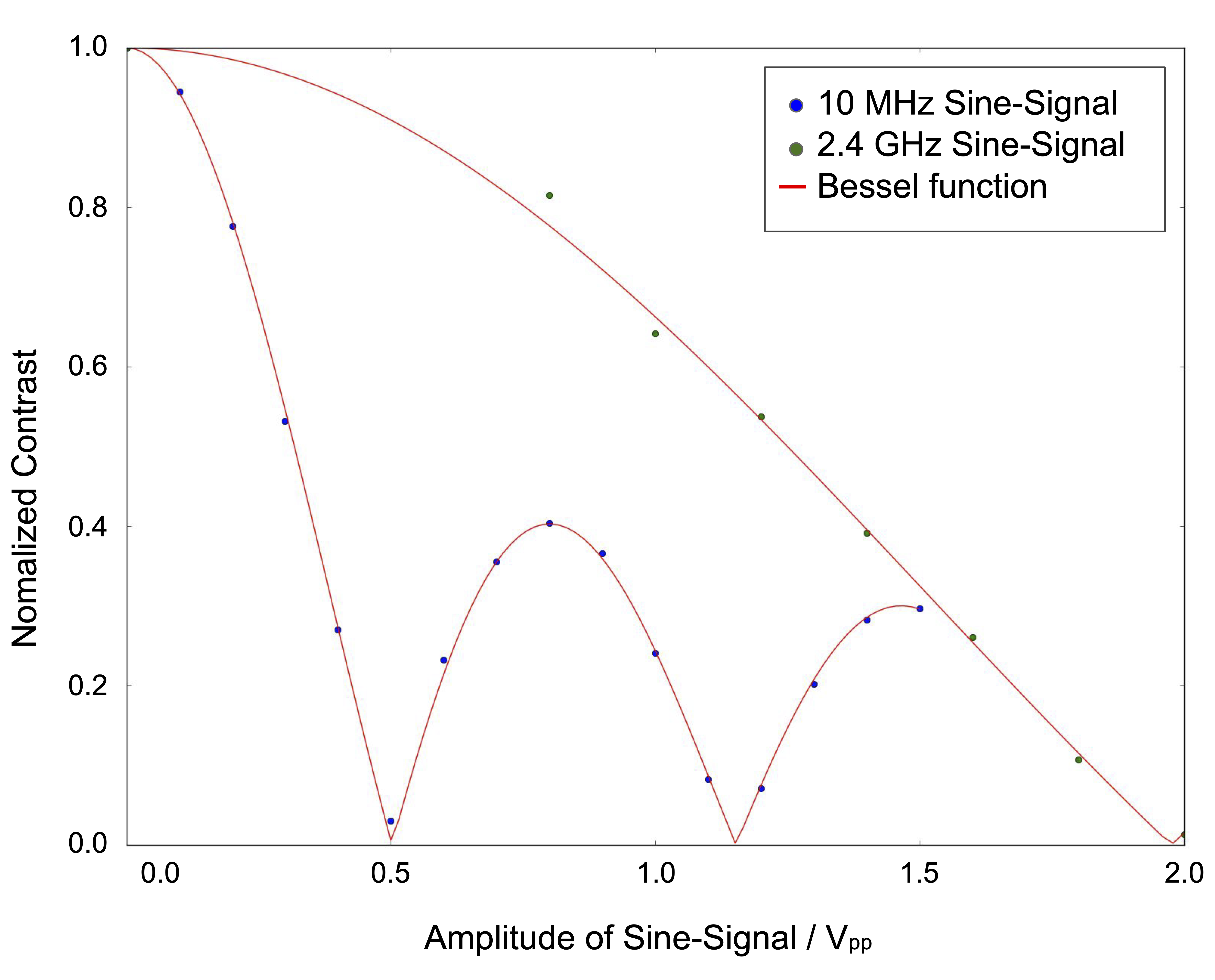}
\par\end{centering}
\caption{Normalized fringe contrast in dependence of amplitude of the sine-signal applied to the electric phase shifter at frequencies of 10~MHz and 2.4~GHz.\label{bessels}}
\end{figure}

Since the biprism by its nature is designed for constant voltages, the question of the RF capability of the biprism as an electric phase shifter is of great interest. A limitation of the temporal resolution for high repetition rates $f = 1/T$ can be the frequency-dependent impedance in combination with the source impedance and possible transmission line effects that distort the signal and lead to an effective phase shift. In order to measure this effective phase shift and to test the switching abilities of the dynamic phase shifter for repetition rates $f$ into the GHz range, the following electron holographic experiment was developed and conducted.\\
\\
If the phase shifter changes the phase difference between both partial waves according to a sine-signal (comparable to [\ref{itm:kralle3}]), i.e.
\[
\varphi(t)=\varphi_{0}\sin(2\pi ft)
\]
and average the interference pattern over a multiple of $T=1/f$,
the interference pattern in (\ref{eq:dynamichologram2}) becomes:
\begin{align}
I_{hol}(\vec{r}) & = 2 + \frac{2\mu}{T} \left[\cos(2\pi\vec{q_{c}}\cdot\vec{r})\int_{0}^{T}\cos(\varphi(t))\,\mathrm{d}t-\sin(2\pi\vec{q_{c}}\cdot\vec{r})\int_{0}^{T}\sin(\varphi(t))\mathrm{d}t\right]\nonumber \\
 & =2 + \frac{2\mu}{T} \left[\cos(2\pi\vec{q_{c}}\cdot\vec{r})\int_{0}^{T}\cos(\varphi_{0}\sin(2\pi ft))\,\mathrm{d}t-\sin(2\pi\vec{q_{c}}\cdot\vec{r})\underbrace{\int_{0}^{T}\sin(\varphi_{0}\sin(2\pi ft))\,\mathrm{d}t}_{=0}\right]\nonumber \\
 & = 2+2\mu\cos(2\pi\vec{q_{c}}\cdot\vec{r})J_{0}(\varphi_{0}). \label{eq:bessel}
\end{align}
As is often the case with frequency modulation in radio communication, the integral expression for the 0-th Bessel function $J_{0}(\varphi_{0})$ was used in (\ref{eq:bessel}). Consequently, the fringe contrast becomes dependent on the amplitude $\varphi_{0}$ of the sinodal phase shifting:
\[
V(\varphi_{0})=\mu\left|J_{0}(\varphi_{0})\right|,
\]
and for the normalized contrast (divided by the contrast of a steady-state hologram $V(0)$)
\begin{equation}
    \frac{V(\varphi_{0})}{V(0)}=\left|J_{0}(\varphi_{0})\right|, \label{eq:bessel2}
\end{equation}
a Bessel function is observed.\\
\\
Fig.~\ref{bessels} shows a measurement of the normalized fringe contrast of recorded electron holograms in dependence of the amplitude (peak to peak-voltage $V_{PP}$) of a sine-signal applied to the electric phase shifter at a frequency of 10~MHz and 2.4~GHz utilizing a IFR 2026 Multi-source Signal Generator. The holograms were recorded under the conditions described in the experimental setup section. The Bessel function $J_{0}$ has its first zero at 2.4048, found at an amplitude of 0.502~$V_{PP}$.\\

\noindent
Here, the voltage of the first zero is determined by a fit of the Bessel function to the measured data. The quality of the fit emphasizes the validity of this approach to measure the effective phase shift. As eq. (\ref{eq:bessel2}) shows, this zero should only depend on the amplitude $\varphi_{0}$ of the sinusoidal signal, not on its frequency. However, due to the impedance of real device like the electric phase shifter, a drop of the effective voltage at the device is expected for higher frequencies $f$. For experiments with frequencies lower than 10~MHz, the zero remains at 0.502~$V_{PP}$.\\
\\
This measurement principle is also applied for a sine-signal with a frequency of 2.4~GHz at the electric phase shifter. The results are shown in Fig.\,\ref{bessels}. Here the normalized contrast drops to zero for a signal amplitude of 1.976~$V_{PP}$. Compared to the measurement at 10~MHz a drop of deflection efficiency to roughly a quarter is observed mostly due to capacitive dampening and reflections in the cabling to the condenser biprism. Therefore, by measuring the zero  amplitudes for different frequencies of the sine-signal, the frequency response of the electric phase shifter (or other devices) can be measured via electron holography.

\section*{Experimental setup}

\begin{figure}[h]
\begin{centering}
\includegraphics[width=1\columnwidth]{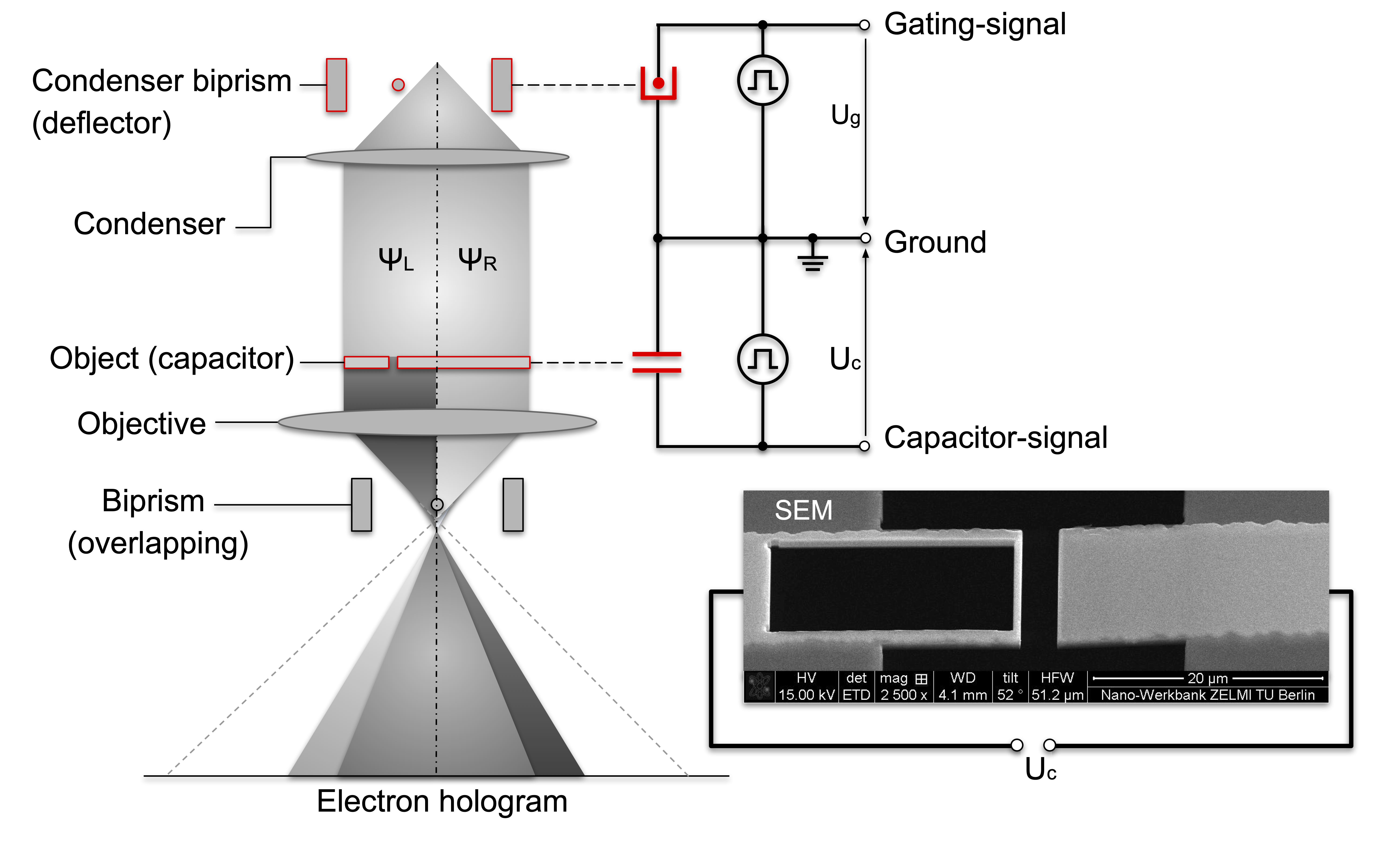}
\par\end{centering}
\caption{Simplified experimental setup with circuit diagram for the electric deflector and the coplanar capacitor as a dynamic phase object.The SEM image of the separated conductor path of a MEMS based heating chip forms the capacitor electrodes. The large rectangular hole in the left  electrode serves as an area for passing the reference wave.\label{fig:setup}}
\end{figure}

The experiments were carried out at the "FEI Titan 80-300 Berlin Holography Special TEM" in an "extended Lorentz holography mode", allowing holography at a large field of view. The electron holograms were recorded at an acceleration voltage of 200~kV using a Gatan US1000 camera with an exposure time of $T_{exp}=$ 4 seconds.\\
\\
The electrical biasing of the specimen was realized utilizing a DENSsolutions Wildfire series S3 heating holder. As a voltage supply for constant voltages, a Keithley 2450 sourcemeter was applied. For dynamic voltages (gating-signal $U_{g}$ and capacitor-signal $U_{c}$ in Fig. \ref{fig:setup}), two arbitrary wave channels of a GW Instek MFG-2260MRA signal generator were used and monitored by a GW Instek GDS-1102B Digital Storage Oscilloscope.

\begin{figure}[h]
\begin{centering}
\includegraphics[width=1\columnwidth]{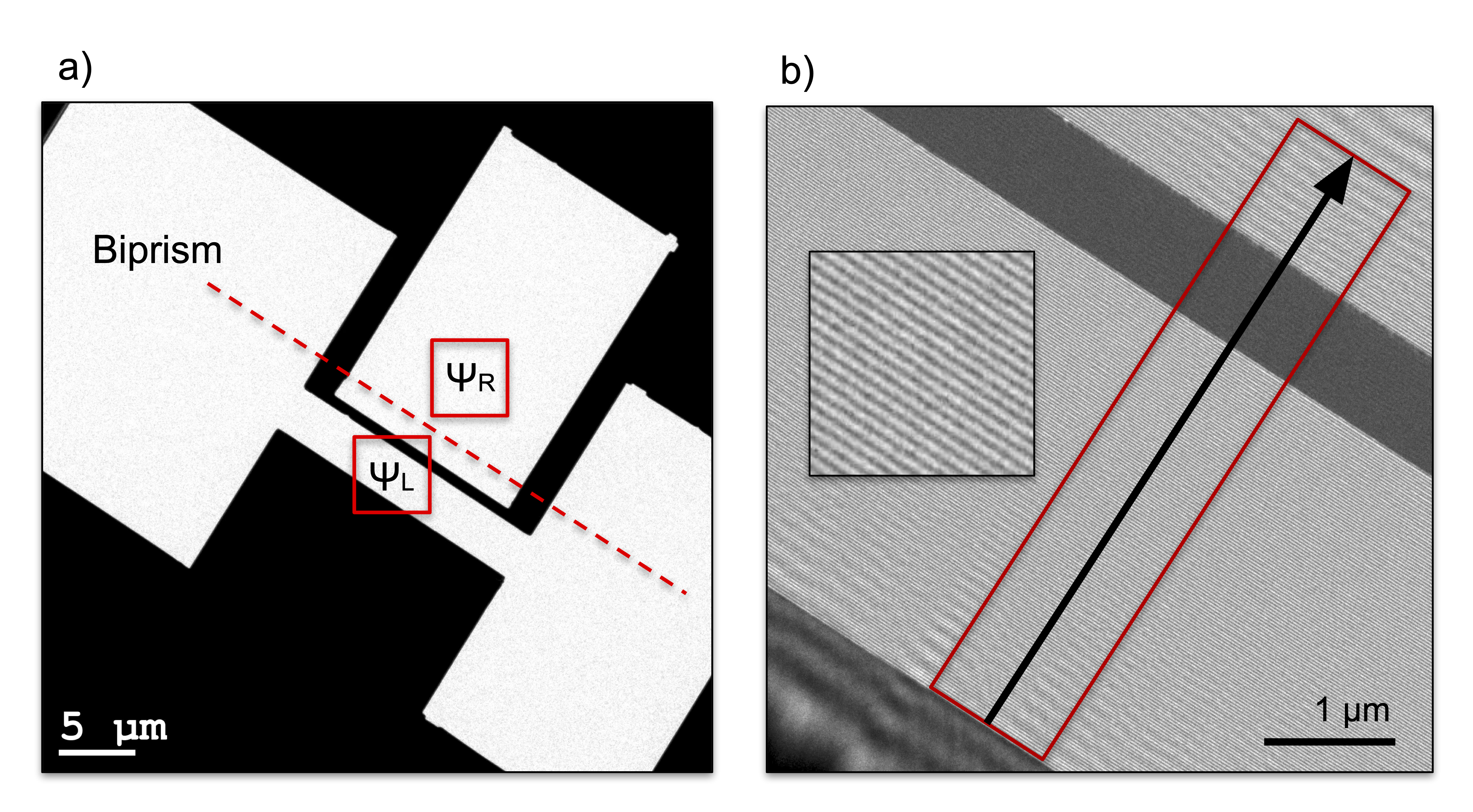}
\par\end{centering}
\caption{a) A TEM overview of the capacitor electrodes and the vacuum regions. The two red squares mark the overlapping areas in the hologram ($\psi_{L}$ and $\psi_{R}$), where as the dashed line the position of the biprism filament. b) Electron hologram of the capacitor with an area zoomed in and the region from which the profile scans for Fig. \ref{fig:capvoltage} are taken from. \label{fig:tem}}
\end{figure}

\noindent
The electric field variation of a periodically switched coplanar capacitor with well known geometric properties was chosen as a demonstrative specimen. Fig. \ref{fig:setup} shows a scanning electron microscopic (SEM) image of the capacitor. The conductor path of a MEMS bases heating chip was separated via focused ion beam (FIB) milling, forming the coplanar electrodes of the capacitor. The large rectangular hole in the left electrode serves as a reference area (reference window), in which the influence of the stray field on the reference wave is drastically reduced. In Fig. \ref{fig:tem} a), a TEM overview of the electrodes and the surrounding vacuum area is shown.\\
\\
In addition, the electrical biasing setup (TEM holder) has been improved with regard to high-frequency suitability. For this purpose, a special multicore coaxial (RG 174) cable was developed and utilized, which enables a wiring according to Fig. \ref{fig:setup} and improves impedance matching.\\
\\
Fig. \ref{fig:setup} shows also a simplified representation of the overall connection to the microscope. The large field of view in the "extended Lorentz holography mode" was achieved by utilizing transfer lenses (TL12, ADL) of image-Cs corrector as a demagnifying objective and placing the crossover close the biprism plane (selective area aperture plane in normal TEM modes). 

\noindent
In this way, hologram fringes with spacing of 24~$nm$ at fields of view up to 10~$\mu m$ can be resolved and distances about 10-15~$\mu m$ can be overlapped at biprism voltages around -60 volts (Fig. \ref{fig:tem} b). The holograms were reconstructed with a Butterworth-like filter of 14th order and a cut off at 14/$\mu m$.

\section*{Results \& discussion}

\begin{figure}[h]
\begin{centering}
\includegraphics[width=1\columnwidth]{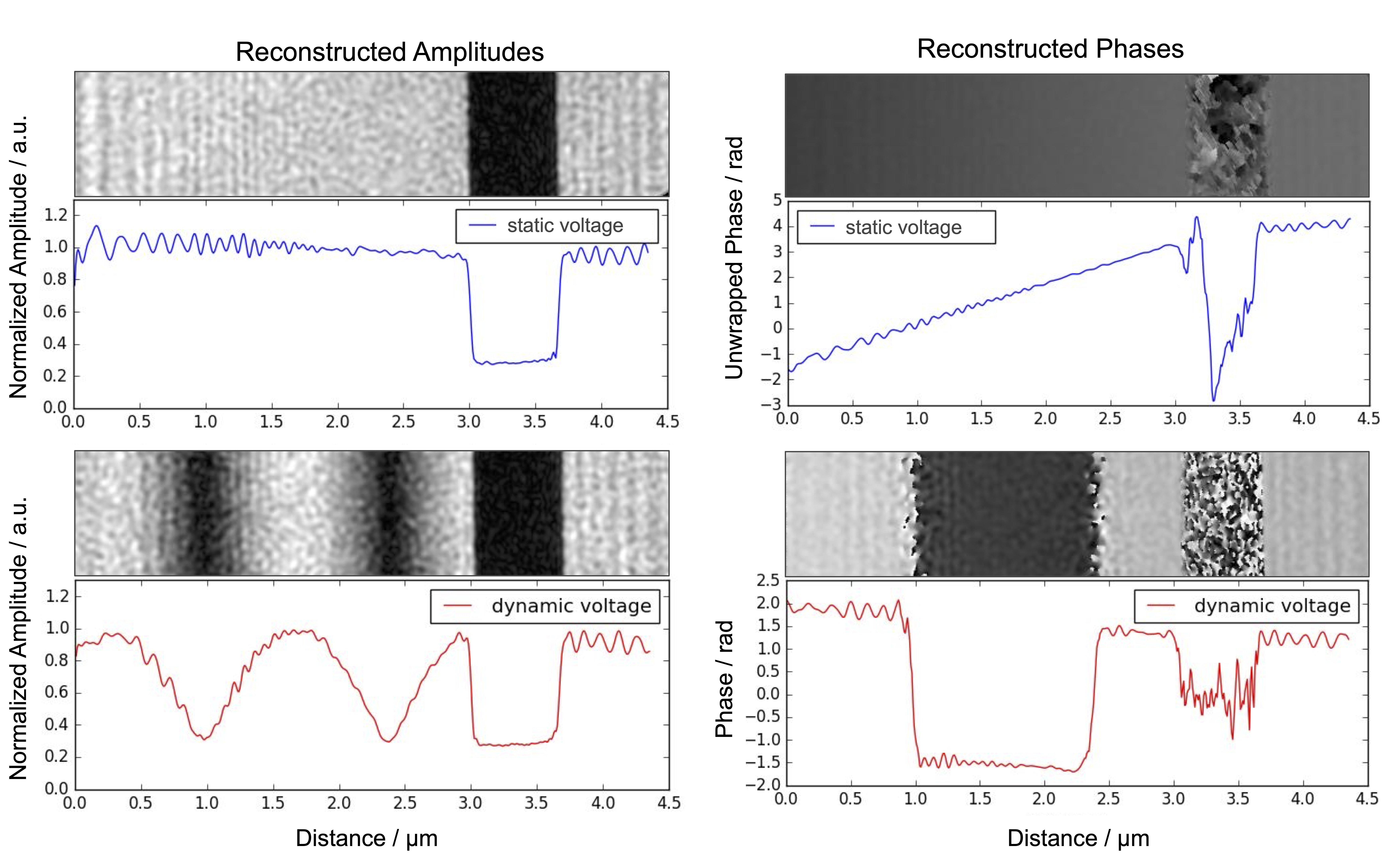}
\par\end{centering}
\caption{Reconstructed amplitudes and phases of conventional holograms recorded according to Fig. \ref{fig:tem} with respective profiles for a static (constant $U_{c}=$ 0.2~$V$ ) and a dynamic voltage (square wave with an amplitude of $U_{c}=$ 0.4~$Vpp$ at a frequency of  $f=$ 1~$MHz$) applied to the capacitor. \label{fig:capvoltage}}
\end{figure}

As described in detail in [\ref{itm:batman}], a voltage applied to the capacitor electrodes causes a phase shift of the electron wave. Fig. \ref{fig:capvoltage} shows cropped amplitudes and phases of reconstructed normalized image waves with their respective profiles. The two separate conventional electron holograms where recorded according to Fig. \ref{fig:tem} with a constant voltage of $U_{c}$ = 0.2~$V$ for the first hologram and an alternating voltage (square wave) with an amplitude of $U_{c}$ = 0.4~$Vpp$ at a frequency of  $f=$ 1~$MHz$ for the second hologram applied to the capacitor. Holograms at the same stage position with short-circuited grounded electrodes are used for normalization. This is crucial for electrical biasing EH as it reduces interfering effects such as specimen charging or setup-related phase shifts thus increasing interpretability.\\
\\
For the static case (constant voltage), the amplitude shows a constant behavior in the vacuum regions (except for Fresnel fringes at the edge regions of the hologram), the unwrapped phase profile shows a linear slope between the capacitor electrodes. This is consistent with previous observations and stems from the proportionality between the projected strongly inhomogeneous electric potential of the biased coplanar capacitor and the modulation of the wavefront. The slope is proportional to the applied voltage $U_{c}$. In this measurement, the left electrode (not visible in Fig. \ref{fig:capvoltage}) is grounded and the right electrode is at a potential of $U_{c}=$ 0.2~$V$. What is remarkable at this point is that the phase behind the electrode (far right in the reference window) is almost constant. This demonstrates that the influence of stray fields on the reference wave can be minimized by such a geometry.

\begin{figure}[h]
\begin{centering}
\includegraphics[width=1\columnwidth]{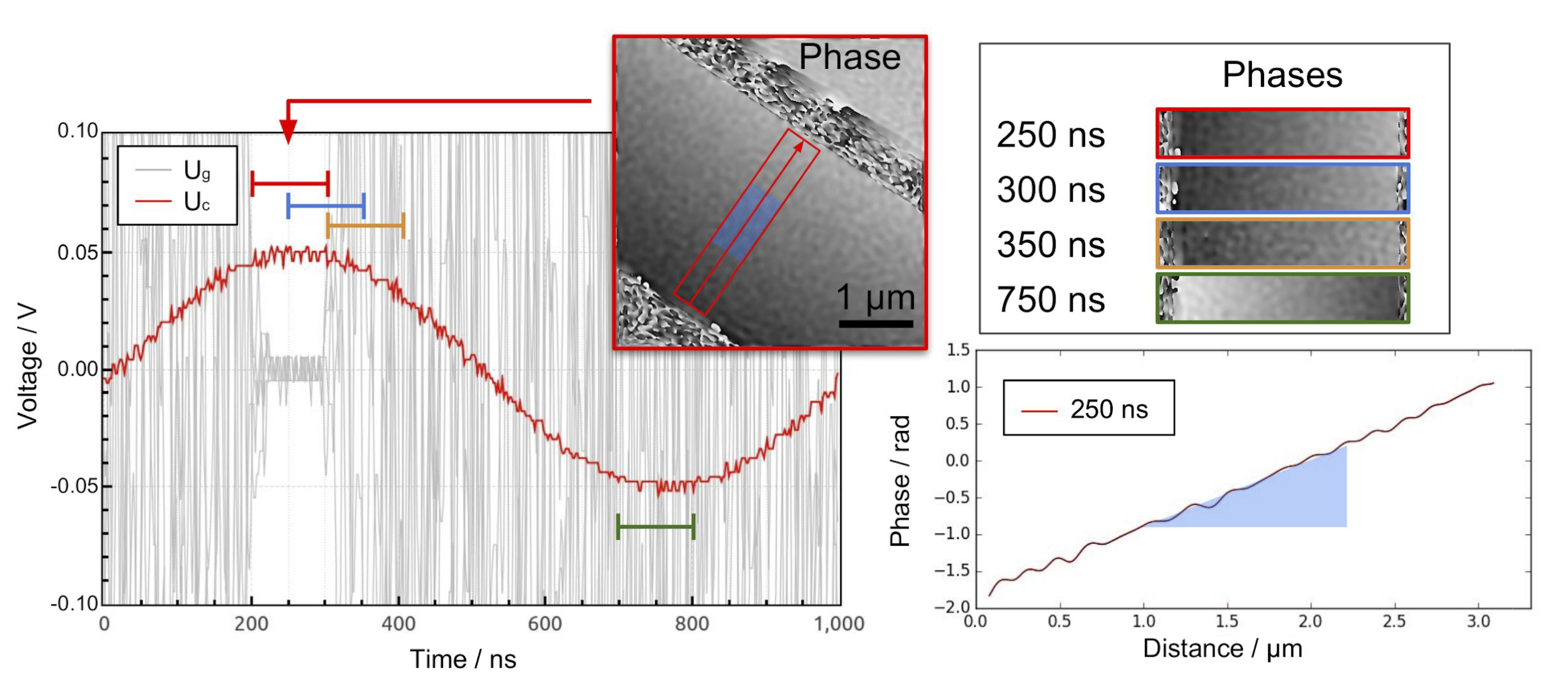}
\par\end{centering}
\caption{Voltage plot: Sine-signal $U_{c}$ applied to capacitor (red line) and the gating signal $U_{g}$ applied to the electrical deflector for the gate position at 250~$ns$ (5 periods are plotted on top of each other). Reconstructed time-resolved phases (100~ns time resolution) for gate positions of 250~$ns$ (red), 300~$ns$ (blue), 350~$ns$ (orange) and 750~ns (green) with an exemplary phase profile illustrating the slope measurement. \label{fig:measure}}
\end{figure}

\noindent
If a dynamic voltage (here a square wave) is applied to the electrodes, a conventional hologram shows only the temporarily averaged interference pattern (double or continuous exposure EH - DEEH [\ref{itm:kralle}]). Then, the reconstructed amplitude of the image wave in Fig. \ref{fig:capvoltage} shows a $|cos(x)|$-like distribution. Whereas the reconstructed phase profile (Fig. \ref{fig:capvoltage}) has a plateau like behavior between the electrodes, which are both typically for DEEH. The projected potential in the reference area also appears to remain constant for alternating voltages. This is particularly advantageous for time-resolved measurements of dynamic samples.\\
\\
Time-resolved electron holography of the periodically changing electric potential of the coplanar capacitor was also performed via iGate. A sine-signal with an amplitude of $U_{c}=$ 0.1~$Vpp$ at a frequency of $f=$ 1~$MHz$ was applied to the capacitor (left electrode remains grounded) and was sampled with a 100~$ns$ gating at 20 equally distributed gate positions (see voltage plot in Fig. \ref{fig:measure}). 

\noindent
The period was oversampled with gate positions every 50~$ns$ (sampling period). At every gate position four EH were acquired and normalized as previously described, reconstructed and averaged [\ref{itm:tore}]. To avoid additional artifacts, the voltage $U_{c}$ was selected so that the phase shift between the electrodes within a measurement interval $\tau$ was less than $2\pi$. Fig. \ref{fig:measure} shows exemplary the reconstructed time-resolved phase for the gate position at 250~ns (maximum of the sine-wave). Additionally, cropped phases in the area between the electrodes are also shown for 300~$ns$, 350~$ns$ and 750~$ns$. Here, it is clearly visible that the slope of the phase is inverted for 750~$ns$ compared to 250~$ns$, since at this gate position a negative voltage is applied to the right electrode. The phase profile for 250~$ns$ in Fig. \ref{fig:measure} serves to illustrate the phase gradient evaluation. The gradient $\frac{d}{dx}\varphi$ perpendicular to the edges of the electrodes is averaged in a central area between them (blue rectangle in the reconstructed time-resolved phase) for each gate position. These averaged slopes should be proportional to the average effective voltage at the capacitor electrodes during the 100~ns gate for each sample.\\

\begin{figure}[h]
\begin{centering}
\includegraphics[width=1\columnwidth]{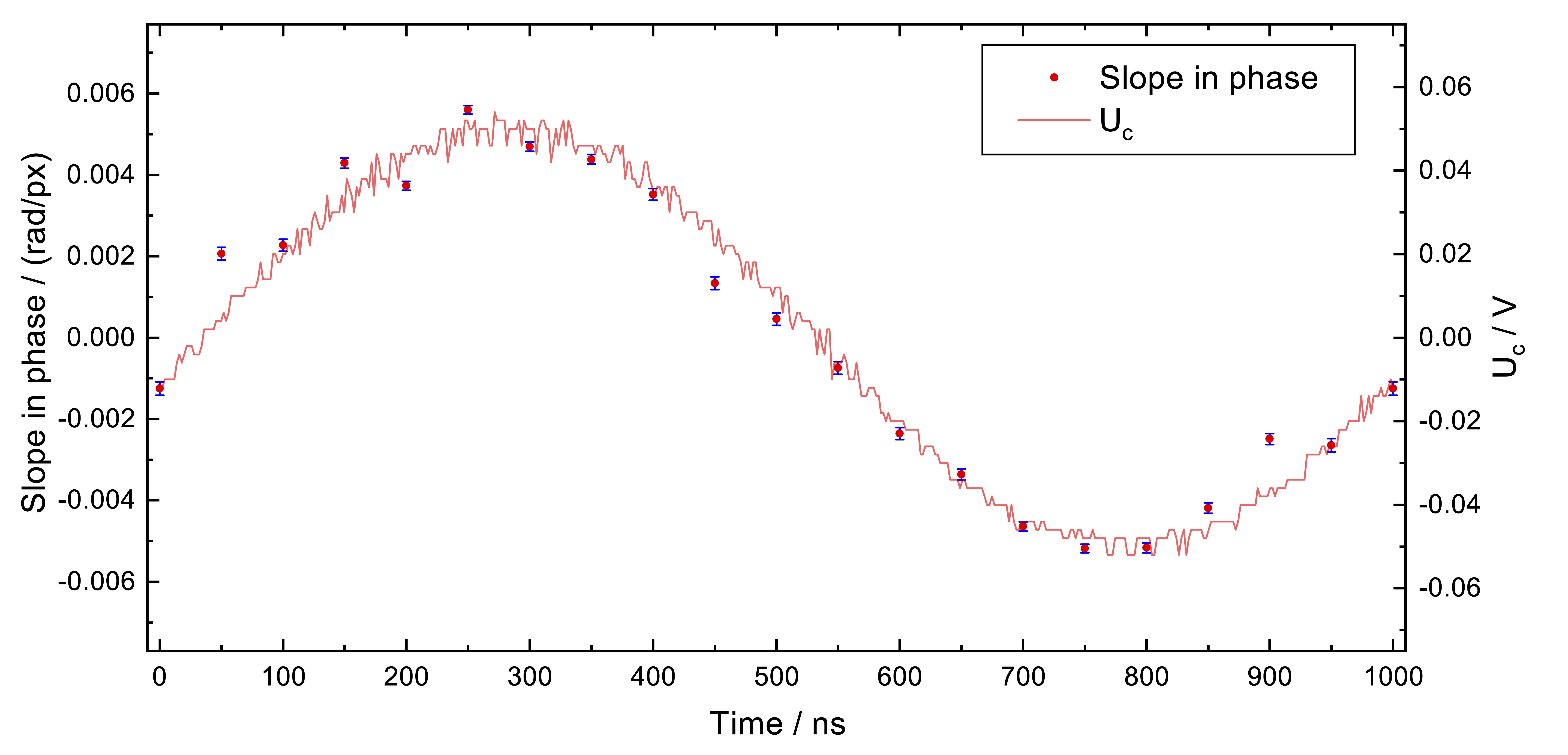}
\par\end{centering}
\caption{Red line: the sine-signal $U_{c}$ applied to the capacitor measured with the oscilloscope. Red dots: the averaged slopes in phase $\frac{d}{dx}\varphi$, measured between the electrodes with 100~ns temporal resolution for 20 samples. \label{fig:sine}}
\end{figure}

\begin{figure}[h]
\begin{centering}
\includegraphics[width=1\columnwidth]{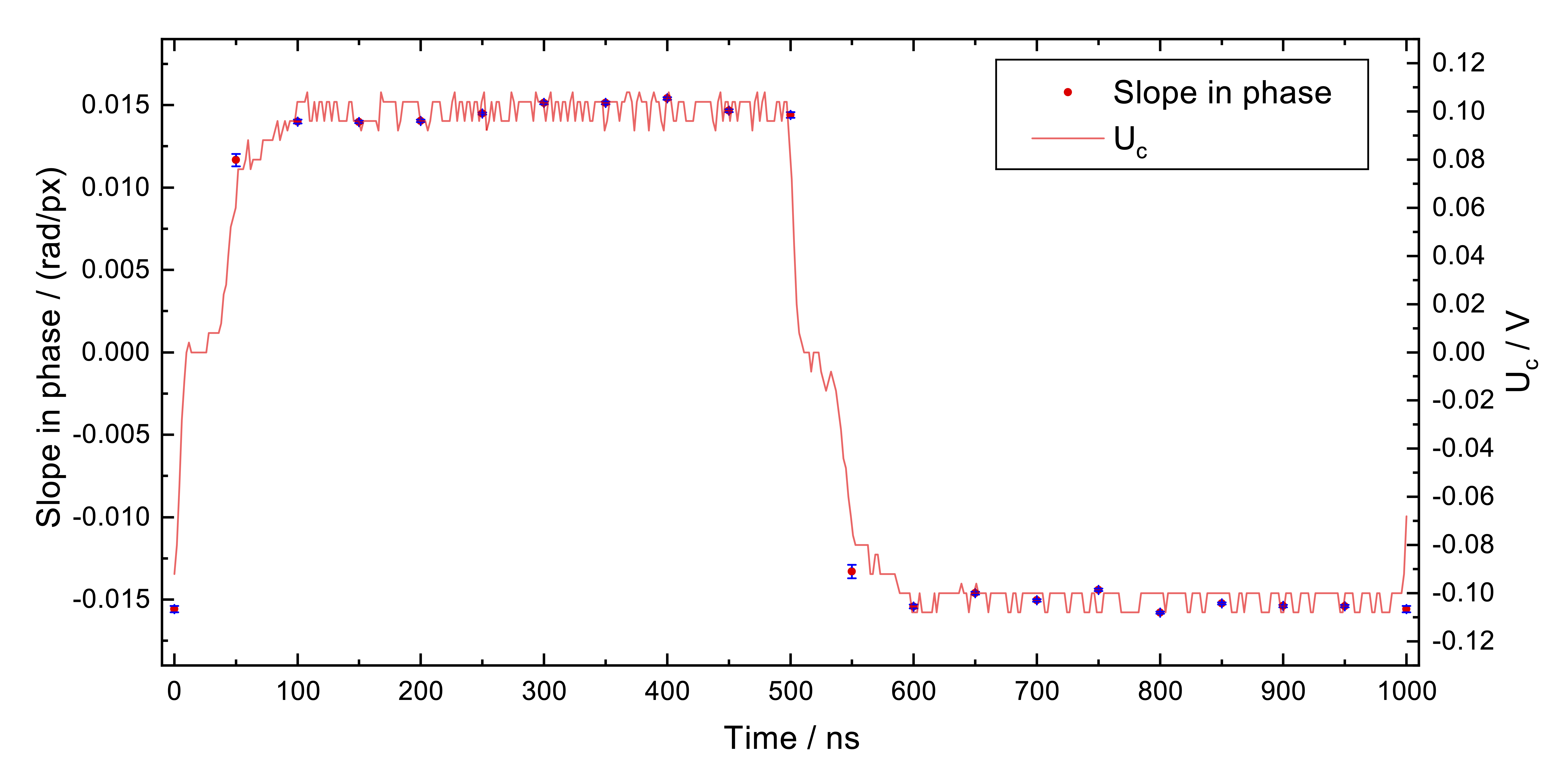}
\par\end{centering}
\caption{Red line: the square-signal $U_{c}$ applied to the capacitor measured with the oscilloscope. Red dots: the averaged slopes in phase $\frac{d}{dx}\varphi$, measured between the electrodes with 100~ns temporal resolution for 20 samples.\label{fig:square}}
\end{figure}

\noindent
Fig. \ref{fig:sine} shows a plot of the averaged slopes of the reconstructed phases for the 20 samples with one period of the applied sine-signal $U_{c}$ (measured with the oscilloscope) in the background. The standard deviation of the slopes within the selected area is displayed as error bars. The plot clearly demonstrates that the phase gradients over the entire period correspond very well with the course of the sine-signal. Slight deviations from 50 to 150~$ns$ are due to instabilities of the setup causing an absolute phase change, which is not corrected in the averaging scheme. The standard deviation of the phase slope (error bars in Fig. \ref{fig:sine}) is relatively small, especially for small voltage changes.\\
\\
The 100 ns measurements show that the capacitor (and the electrical biasing setup) reproduces the applied sine-signal without any notable changes in amplitude and phase even at 1~MHz (compared to measurements at lower frequencies which were carried out separately).\\
\\
For higher frequencies (in the range of tens of MHz), a voltage drop at the electrodes is expected due to the capacitive damping. This effective voltage $U_{c-eff}$ at the electrodes leads to a change in phase slope, which would make time-resolved EH suitable for a frequency response analysis. At the moment, however, the time resolution of our method is limited to 100 $ns$ due to bandwidth limitation of the signal generator.\\
\\
For signals with higher harmonics, effects of capacitive dampening should already be seen for lower frequencies. Fig. \ref{fig:square} shows the results for an analogous measurement of a 1~$MHz$ square-signal with an amplitude of $U_{c} =$ 0.2~$Vpp$. Also here, the averaged gradients in the measured phase reflect the applied square wave signal very well. Among other things, the deviations of the red curve from the ideal rectangular shape are due to reflections of the electrical signal at cable transitions within the biasing holder (impedance losses). Especially the step in the voltage transition at 25~$ns$ and 525~$ns$ indicates such signal reflections. The general signal curve also shows the typical charging and discharging behaviour of a capacitor.

\noindent
Also for the square-signal, the measured gradients reflect the voltage curve very well. The first gradients after the voltage reversal at 50 ns and 550 ns show a smaller value than the other slopes. On such short time scales, an increasing behaviour of the effective voltage $U_{c-eff}$ at the electrodes after a polarity reversal (rise time $\tau_r$) is well compatible with the theoretical description of an RC circuit. The larger standard deviations of the slopes (error bars in Fig. \ref{fig:square}) at these sections are due to DEEH effects, which lead to noisy regions in the reconstructed phase. This measurement shows that time-resolved EH principally allows the direct measurement of the rise time $\tau_r$ of such or similar systems directly at the location of the event. Additional capacitance measurements (Keysight E4980A LCR meter) showed, however, that the rise time of this system with $\tau_r$ = 150~ns is in the range of the current time resolution, so that an exact measurement of $\tau_r$ using iGate is currently not possible.\\
\\
Obviously, the time resolution $\tau$ depends here on bandwidth limitations of the signal generator and impedance losses at the deflection system or the biasing holder. To a certain extent, these problems can be compensated by forward correction of the control signals. In particular, a slight amplitude modulation of the gating-signal (by smoothing the gate edges) reduces Gibbs-like-artifacts significantly. As a result, significantly higher time resolutions can be achieved with the same signal generator.

\section*{Conclusion \& outlook}

Time-resolved electron holographic measurements of periodic signals in the MHz range using interference gating are demonstrated. By means of an electric phase shifter in the condenser system, a time resolution $\tau$ in the nanosecond range is achieved. The method allows measuring the capacitive damping of an electrically biased specimen. Currently, the time resolution $\tau$ is foremost limited by the bandwidth limitations of the signal generator and the electrical biasing system.\\
\\
An investigation of the phase shifter itself shows that iGate can also be extended into the picosecond range. Hurdles, such as the bandwidth limitation of the signal generator or the biasing holder, can be overcome by using suitable GHz-electronics and reasonable frequency matching. The use of dedicated hardware, such as fast beam shutters based on RF cavities [\ref{itm:rens}], could also permit significantly faster switching times and thus time resolutions. Furthermore, it is conceivable that the method could also be used in conjunction with other phase modulation applications. Programmable phase plates [\ref{itm:ADAPTEM}] for instance can thus be used to modulate the phase of the incident electron wave in space and time.\\
\\
Our current results clearly show the advantages of this method in the field of dynamic potentiometry of electric and magnetic fields. By accessing the potential distribution inside and outside of a sample in combination with very high spatial and time resolution, the interference-gating paves the way for a more precise observation of fast dynamic processes. Although the transition to more complex samples (e.g. electrically connected semiconductor devices) is a challenge in terms of preparation, iGate would provide insights into the dynamic switching behavior (e.g. growth and shrinkage of space charge regions). In the case of a silicon diode with medium doping concentrations, a spatial resolution of 20 $nm$ and a time resolution of around 100 ns should be sufficient to investigate switching effects in the space charge region. For improvement of the setup, a frequency matching of the electrical biasing system is necessary. Overall, from a strictly physical point of view, there is no reason why the interference gating cannot also be implemented in the picosecond or femtosecond range, which further expands the possible field of application. 

\section*{Acknowledgements}

The authors thank Dr. Dirk Berger for his help in the development and the FIB fabrication of the coplanar capacitor with reference window. The authors are grateful to Dr. Gunter Larisch for providing the IFR 2026 multi-source signal generator and Norman Susilo for the capacitance measurements.

\section*{References}

\begin{enumerate}
    
  \item O. Bostanjoglo, R. Elschner, Z. Mao, T. Nink, and M. Weingärtner, Ultramicroscopy 81 (2000) 141. \label{itm:bost1}

 \item A.H. Zewail, Science 328 (2010) 187. \label{itm:zewail}
 
 \item A. Feist, N. Bach, N.R.d. Silva, T. Danz, M. Möller, K.E. Priebe, T. Domröse, J.G. Gatzmann, S. Rost, J. Schauss, S. Strauch, R. Bormann, M. Sivis, S. Schäfer, C. Ropers, Ultramicroscopy 176 (2017) 63. \label{itm:feist}
 
 \item F. Pitters et al 2019 JINST14 P05022. \label{itm:pitters}

 \item H. Wahl, Bildebenenholographie mit Elektronen (Habilitationschrift, Universität Tübingen, 1975). \label{itm:kralle}

 \item C. Gatel , F. Houdellier , E. Snoeck , J. Phys. D 49 (2016) 324001. \label{itm:kralle2}

 \item V. Migunov, C. Dwyer, C.B. Boothroyd, G. Pozzi, R.E. Dunin-Borkowski, Ultramicroscopy 178 (2017) 48. \label{itm:kralle3}
 
  \item K. Soma, S. Konings, R. Aso, N. Kamiuchi, G. Kobayashi, H. Yoshida, S. Takeda, Ultramicroscopy 181 (2017) 27. \label{itm:soma}

   \item F. Houdellier, G.M. Caruso, S. Weber, M.J. Hÿtch, C. Gatel, A. Arbouet, Ultramicroscopy (2019), doi: https://doi.org/10.1016/j.ultramic.2019.03.016 \label{itm:flor}
 
   \item Twitchett, A.C., Dunin-Borkowski, R.E. and Midgley, P.A., 2002. Quantitative electron holography of biased semiconductor devices. Physical review letters, 88(23), p.238302. \label{itm:twitchett1}
  
  \item A.C. Twitchett, R.E. Dunin-Borkowski, R.J. Hallifax, R.F. Broom, and P.A. Midgley, Microscopy and Microanalysis 11 (2005) 66. \label{itm:twitchett2}

  \item Sato, Y., Hirayama, T. and Ikuhara, Y., Physical review letters 107 (2018), p.187601. \label{itm:electric1}
 
  \item Winkler, C. R., Damodaran, A. R., Karthik, J., Martin, L. W., Taheri, M. L. (2012). Direct observation of ferroelectric domain switching in varying electric field regimes using in situ TEM. Micron, 43(11), 1121-1126. \label{itm:electric2}

   \item Gueye, I., Le Rhun, G., Renault, O., Cooper, D., Defay, E., Barrett, N. (2018). Correlation of electrical characteristics with interface chemistry and structure in Pt/Ru/PbZr0. 52Ti0. 48O3/Pt capacitors after post metallization annealing. Applied Physics Letters, 113(13), 132901. \label{itm:coop1}

  \item Haas, B., Rouvière, J. L., Boureau, V., Berthier, R., Cooper, D. , Ultramicroscopy 198, (2018) 58-72. \label{itm:coop2}
 
  \item  T. Niermann, M. Lehmann, T. Wagner, Ultramicroscopy 182 (2017) 54-61. \label{itm:batman}
  
  \item H. Lichte and M. Lehmann, Rep. Prog. Phys.71 (2008) 016102 \label{itm:lichte}
  
  \item T. Niermann, M. Lehmann, Micron 63 (2014) 28. \label{itm:tore}

  \item Q. Ru, J. Endo, T. Tanji, A. Tonomura,  Applied Physics Letters 59 (1991) 2372–2374. \label{itm:ru1}
    
  \item Q. Ru, G. Lai, K. Aoyama, J. Endo, A. Tonomura,  Ultramicroscopy 55 (1994) 209–220. \label{itm:ru2}
    
  \item van Rens, J.F.M. , Verhoeven, W. , Kieft, E. R., Mutsaers, P. H.A. , Luiten, O. J., Applied Physics Letters. 2018 ; Vol. 113, No. 16. \label{itm:rens}

  \item  J. Verbeeck, A. Beche, K. Muller-Caspary, G. Guzzinati, M.A. Luong, M. Den Hertog, Ultramicroscopy 190 (2018) 58-65. \label{itm:ADAPTEM}

\end{enumerate}

\end{document}